\documentclass[global,twocolumn]{svjour}
\usepackage{graphics}
\journalname{myjournal}
\begin{document}
\title{Ultrafast Optical Spectroscopy of Micelle-Suspended
Single-Walled Carbon Nanotubes}
\author{J. Kono\inst{1}\thanks{Corresponding author; kono@rice.edu.}
\and G. N. Ostojic\inst{1} \and S. Zaric\inst{1}
\and M. S. Strano\inst{2}\thanks{Present address: Department of Chemical and
Biomolecular Engineering, University of Illinois, Urbana, IL.}
\and V. C. Moore\inst{2} \and J. Shaver\inst{2}
\and \\ R. H. Hauge\inst{2} \and and R. E. Smalley\inst{2}
}                     
%
%
\institute{Department of Electrical and Computer Engineering,
Rice Quantum Institute, and Center for Nanoscale Science and Technology,
Rice University, Houston, Texas 77005 \and
Department of Chemistry,
Rice Quantum Institute, and Center for Nanoscale Science and Technology,
Rice University, Houston, Texas 77005}
\date{Received: date / Revised version: date}
%
\maketitle
\begin{abstract}

We present results of wavelength-dependent ultrafast pump-probe experiments
on micelle-suspended single-walled carbon nanotubes.
The linear absorption and photoluminescence spectra of the samples
show a number of chirality-dependent peaks, and consequently,
the pump-probe results sensitively depend on the wavelength.
In the wavelength range corresponding to the second van Hove singularities (VHSs),
we observe sub-picosecond decays,
as has been seen in previous pump-probe studies.  We ascribe these ultrafast
decays to intraband carrier relaxation.
On the other hand, in the wavelength range corresponding to the first VHSs,
we observe two distinct regimes in ultrafast carrier relaxation:
fast (0.3-1.2 ps) and slow (5-20 ps).
The slow component, which has
not been observed previously,
is resonantly enhanced whenever the pump photon energy
resonates with an interband absorption peak, and we attribute it to
radiative carrier recombination.
Finally, the slow component is dependent on the pH
of the solution, which suggests an important role
played by H$^+$ ions surrounding the nanotubes.

\end{abstract}
\section{Introduction}
\label{intro}

Single-walled carbon nanotubes (SWNTs) provide a variety of new
opportunities for the exploration of one-dimensional (1-D) quantum physics
as well as novel device applications.
Although their mechanical and electrical properties have been extensively
studied during the past decade, there has been only limited success in exploring
their optical, magnetic, and magneto-optical properties.
In addition to what is commonly expected for excitons in 1-D systems
\cite{Loudon59AJP,ElliotLoudon59JPCS,OgawaTakagahara91PRBRC},
SWNTs are expected to show a new class of
optical phenomena that
arise from their unique tubular structure with varying diameters and chiral angles.
Linear and nonlinear optical coefficients are expected to be
diameter- and chirality-dependent, and an external magnetic field is
expected to induce drastic modifications on
their band structure via the Aharonov-Bohm phase
\cite{AjikiAndo93JPSJ,AjikiAndo93BJPSJ,TianDatta94PRB,Lu95PRL}.
Furthermore, nonlinear harmonic generation is expected to be highly
selective for creating certain orders of high harmonics \cite{AlonetAl00PRL}.

However, these predicted optical phenomena and properties
have not been probed experimentally.
This is because in standard production methods SWNTs
appear in the form of bundles (or `ropes')
due to their strong van der Waals forces.  This `roping' results in significant broadening
of electronic states, smearing out any chirality-dependent features in optical spectra
\cite{Katauraetal99SM,Ichidaetal99JPSJ,Kazaouietal99PRB,Ugawaetal99PRB,Kazaouietal00PRB,Hwangetal00PRB,Saitoetal01Book}.
Very recently, a new technique for producing individually suspended SWNTs
has been reported \cite{oconnelletal02Science}.
These samples revealed, for the first time, a number of
clearly observable peaks in linear absorption and photoluminescence (PL) spectra,
corresponding to interband transitions in different types of tubes.
A subsequent PL excitation (PLE) spectroscopy study successfully
provided detailed peak assignments \cite{BachiloetAl02Science,WeismanBachilo03NL}.
These seminal studies have provided opportunities to systematically study
other predicted, unique optical properties of SWNTs.

Ultrafast laser spectroscopy is one of the best methods for
measuring carrier distribution functions and relaxation mechanisms.
Using a wide variety of techniques employing femtosecond pulses, one can directly
examine dynamical processes after creation of electron-hole pairs across the band gap,
i.e., carrier dephasing
processes, carrier-carrier scattering, carrier-phonon scattering,
exciton formation, and radiative and non-radiative recombination dynamics.
There have been several ultrafast optical studies of
SWNTs \cite{HertelMoos00PRL,ChenetAl02APL,HanetAl02APL,LauretetAl03PRL}.
Hertel and Moos \cite{HertelMoos00PRL}
used ultrafast photoemission spectroscopy to determine
thermalization times in metallic SWNTs to be $\sim$200 fs.
Three groups \cite{ChenetAl02APL,HanetAl02APL,LauretetAl03PRL} independently observed
subpicosecond to a ps lifetimes for semiconducting SWNTs in pump-probe
spectroscopy.  Although such ultrashort lifetimes can be expected for {\em intraband}
relaxation towards the band edge, the microscopic origin of the fast
{\it interband} decay (which is likely to be non-radiative) is still unclear.
One important issue is that these studies were performed
on {\it bundled} SWNTs, which did not show any PL.
It is thus desired
to carry out ultrafast spectroscopy on SWNT samples that show chirality-assigned
absorption and PL peaks.

In this article, we report results of
a femtosecond pump-probe study of chirality-assigned SWNTs.
The linear absorption and PL spectra of such
micelle-suspended SWNT samples
show a number of chirality-dependent peaks, and
as a result,
the pump-probe data sensitively depends on the wavelength used.
In the wavelength range corresponding to the second van Hove singularities (VHSs),
we observe ultrafast (sub-picosecond) decays,
as has been reported previously \cite{ChenetAl02APL,HanetAl02APL,LauretetAl03PRL}.
We ascribe these ultrafast
decays to intraband carrier relaxation.
On the other hand, in the wavelength range corresponding to the first VHSs,
we observe {\it two distinct regimes} in ultrafast carrier relaxation:
fast (0.3-1.2 ps) and slow (5-20 ps).
The slow component has not been observed previously,
is resonantly enhanced when the pump photon energy
matches an absorption peak, and is attributed to radiative carrier recombination.
The slow component is also strongly dependent on the pH
of the solution, especially in large diameter tubes.

\section{Samples Studied}
\label{samples}

\begin{figure}
\begin{center}
\resizebox{0.37\textwidth}{!}{%
\includegraphics{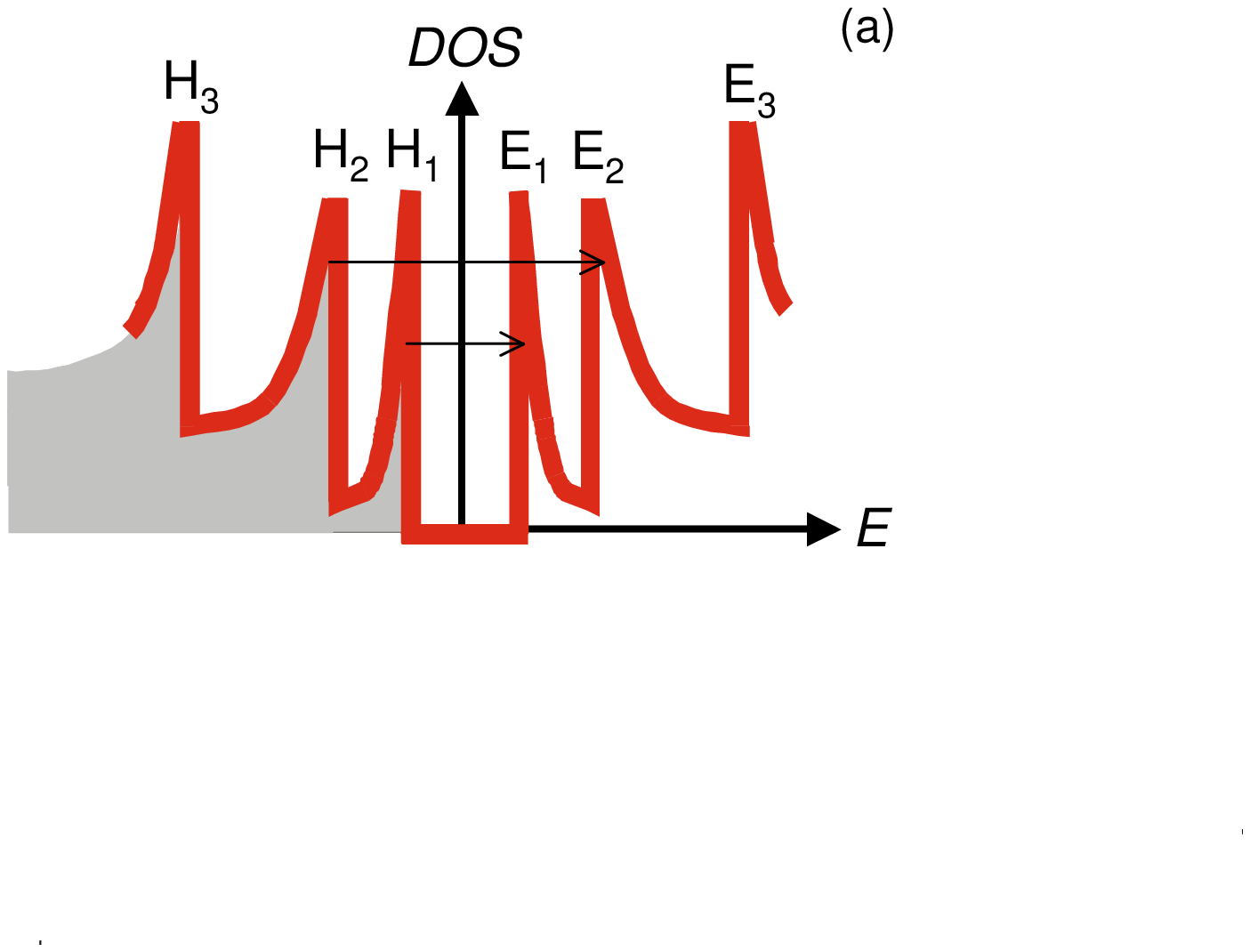}
}
\resizebox{0.49\textwidth}{!}{%
\includegraphics{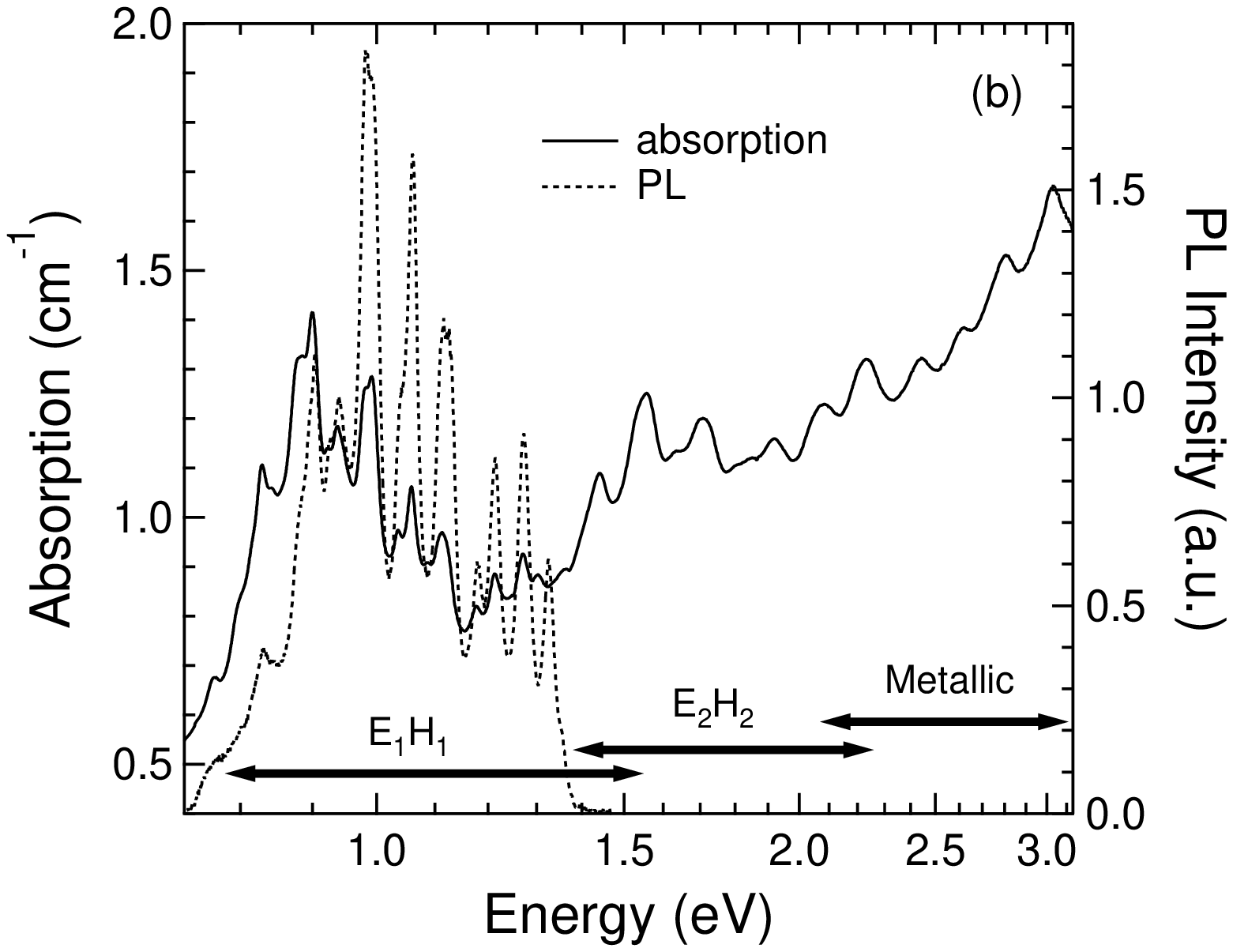}
}
\caption{(a) Schematic density of states vs. energy for a semiconducting
carbon nanotube.  (b) Typical linear absorption (solid line) and
photoluminescence (dashed line) spectra of SDS-suspended
single-walled carbon nanotubes in heavy water (ph = 9) at room temperature.
The photoluminescence was excited at 733 nm.}
\label{linear}
\end{center}
\end{figure}

The SWNTs studied in the present work were dispersed in aqueous sodium dodecyl
sulfate (SDS) surfactant, sonicated, and centrifuged, which left micelle-suspended
nanotube solutions.  Details of the sample preparation method were described
previously \cite{oconnelletal02Science}.

Typical linear absorption and photoluminescence (PL) spectra for such
micelle-suspended SWNTs are shown in Fig. \ref{linear}(b), together with a schematic
density of states versus energy of semiconducting SWCNTs in Fig. \ref{linear}(a).
The PL peaks occur in the near-infrared ($\sim$0.9-1.4 eV) and are due to transitions
involving the first conduction (E$_1$) and valence (H$_1$) subbands.
The absorption spectrum shows peaks from the near-infrared to the visible range,
consisting of three overlapping bands: E$_1$H$_1$ transitions in semiconducting
tubes (0.78-1.55 eV), E$_2$H$_2$ transitions in semiconducting tubes (1.38-2.26 eV),
and transitions in metallic tubes (2.07-3.11 eV).
Detailed analyses and interpretations of these linear absorption/emission features
are described in \cite{BachiloetAl02Science,WeismanBachilo03NL}.


%
%
\section{Experimental Setup}
\label{setup}

We performed wavelength-dependent, degenerate pump-probe measurements
using $\sim$150 fs pulses from an optical parametric amplifier (OPA)
pumped by a chirped pulse amplifier (Clark-MXR CPA2010).  We used a low
pulse-repetition rate (1 kHz) for
minimizing the average power and reducing any thermal effects
while keeping the fluence high.  To detect small photoinduced changes in
probe transmission, we synchronously chopped the pump beam at 500 Hz and
measured the transmission with ($T$) and without ($T_{0}$)
the pump using two different gates of a box-car integrator.
The smallest detectable differential transmission
[$(T-T_0)/T_0 \equiv \Delta T / T_0$] was $\sim$10$^{-4}$.
We used a noncollinear
geometry with a pump beam diameter of $\sim$220 $\mu$m in the
overlap area.  We tuned the
OPA throughout the range of first subband transitions (i.e., E$_1$H$_1$
transitions; see Fig. \ref{linear}) with photon
energies $h\nu$ = 0.8 eV to 1.13 eV
(wavelengths $\lambda$ = 1.1 $\mu$m to 1.55 $\mu$m).  In addition, by directly
using the CPA beam ($h\nu$ = 1.60 eV, $\lambda$ = 775 nm),
we probed a region of second subband transitions (E$_2$H$_2$).

\section{Experimental Results}
\label{results}

\subsection{First vs. Second Subband Excitations}
\label{typical-data}

\begin{figure}
\begin{center}
\resizebox{0.48\textwidth}{!}{%
\includegraphics{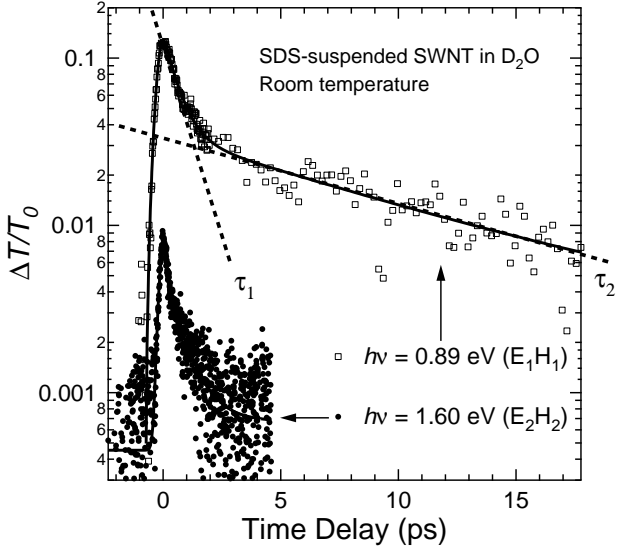}
}
\caption{Ultrafast pump-probe data at two wavelengths,
corresponding to first and second subband transitions.
A very fast single decay is seen in the second subband case while two decay
components ($\tau_1$ and $\tau_2$) are clearly seen in the first subband case.
Solid lines show Gaussian and exponential fits in the appropriate delay regimes.
}
\label{typical}
\end{center}
\end{figure}

Figure \ref{typical} shows typical differential
transmission ($\Delta T /T_0$) data as a function of time delay.  Two traces are
shown, taken at 0.89 eV (1393 nm) and 1.60 eV (775 nm),
corresponding to E$_1$H$_1$ and
E$_2$H$_2$ ranges, respectively (refer to Fig. \ref{linear}).
Both show a positive change (or an increase) in
transmission, i.e., photoinduced bleaching, which is consistent with
band filling.  An exponential fit reveals a fast, single decay time of 770 fs for the
E$_2$H$_2$ transition, which is consistent with earlier reports and
can be explained by the very fast
intraband carrier relaxation towards the band edge \cite{LauretetAl03PRL}.
On the contrary, data in the range of
E$_1$H$_1$ transitions exhibit {\em multiple exponential decays}.  The major
decay of the photoinduced signal happens in the first picosecond (with
decay time $\tau_1$), which is followed by a much slower relaxation process.
For the particular data shown in Fig. \ref{typical},
we obtained an exponential decay time of $\tau_2$ $\approx$ 10 ps.  This long
decay time has not been reported previously for either metallic \cite{HertelMoos00PRL}
or semiconducting SWNTs \cite{ChenetAl02APL,HanetAl02APL,LauretetAl03PRL}.

\subsection{Pump Fluence Dependence}
\label{fluence-dep}

\begin{figure}
\resizebox{0.46\textwidth}{!}{%
\includegraphics{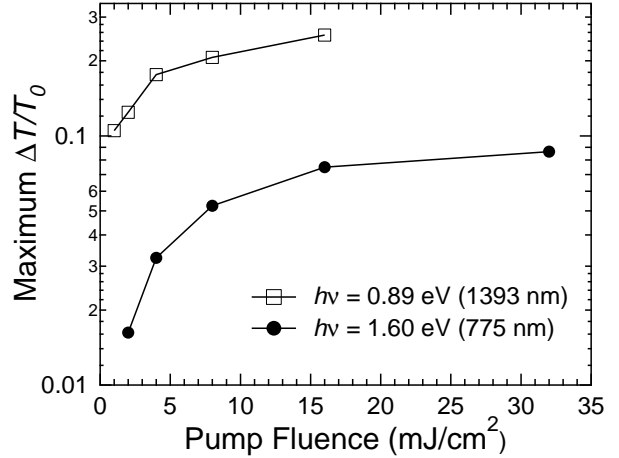}
}
\caption{The pump fluence dependence of the maximum photoinduced transmission change
at the two wavelengths corresponding to the two traces in Fig. \ref{typical}.
In both cases clear saturation is seen.
}
\label{saturation}
\end{figure}

\begin{figure}
\begin{center}
\resizebox{0.48\textwidth}{!}{%
\includegraphics{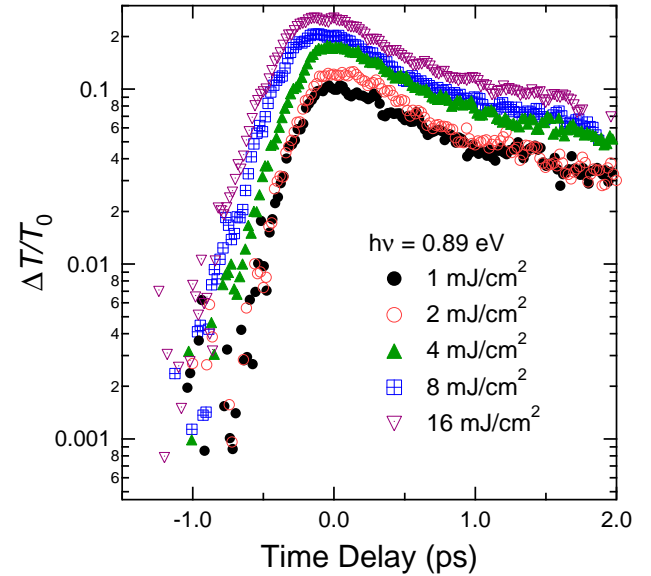}
}
\caption{Power-dependent pump-probe spectroscopy data taken at
a wavelength of 1402 nm (or a photon energy of 0.884 eV).
The decay dynamics do not show any power dependence, excluding
any nonlinear recombination mechanisms.}
\label{power}
\end{center}
\end{figure}

For both first and second subband transitions, the pump fluence
dependence of the maximum value of $\Delta T/T_0$ reveals clear
saturation at high fluences, as shown in Fig. \ref{saturation}.  This implies
that, in the saturation regime, most of the carrier states are filled up and
thus the sample absorption is nearly completely quenched.  A careful
analysis of the differential transmission decays for the $h\nu$ = 0.89 eV case
showed that relaxation
dynamics are not dependent on the pump fluence, including the saturation
regime.
As an example, in Fig. \ref{power} we show pump-probe data at a wavelength of 1402 nm
for four different pump fluences.
This precludes the possibility of any nonlinear recombination process
such as the Auger recombination.

\subsection{Resonant vs. Non-resonant Excitations}
\label{peak-valley}

\begin{figure}
\begin{center}
\resizebox{0.50\textwidth}{!}{%
\includegraphics{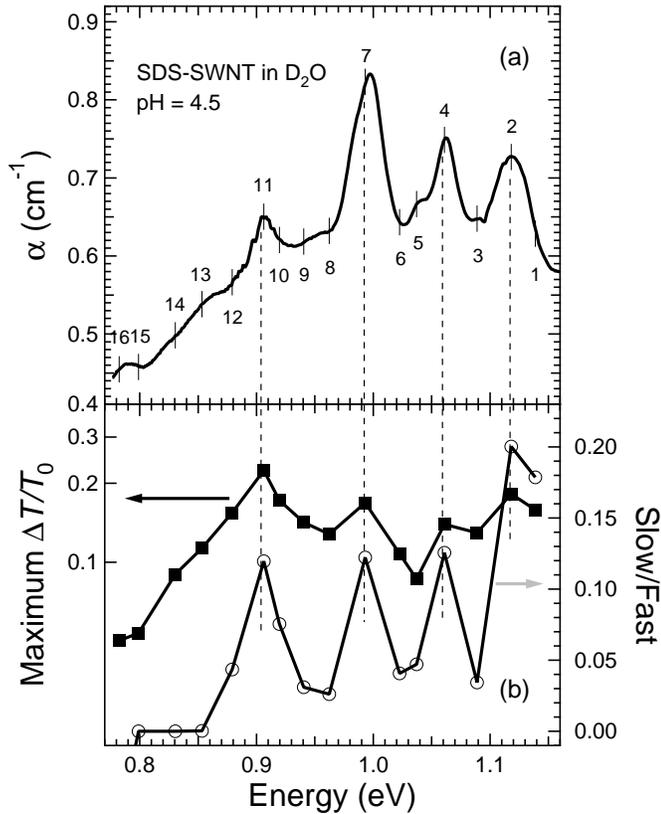}
}
\caption{(a) Linear absorption in the E$_1$H$_1$ range.  The numbers (1-16)
correspond to the 16 photon energies at which pump-probe measurements
were made.  (b) The peak value of $\Delta T/T_0$ (left axis) and the
ratio of slow to fast components (right axis) vs. photon
energy.}
\label{summary}
\end{center}
\end{figure}

\begin{figure}
\begin{center}
\resizebox{0.48\textwidth}{!}{%
\includegraphics{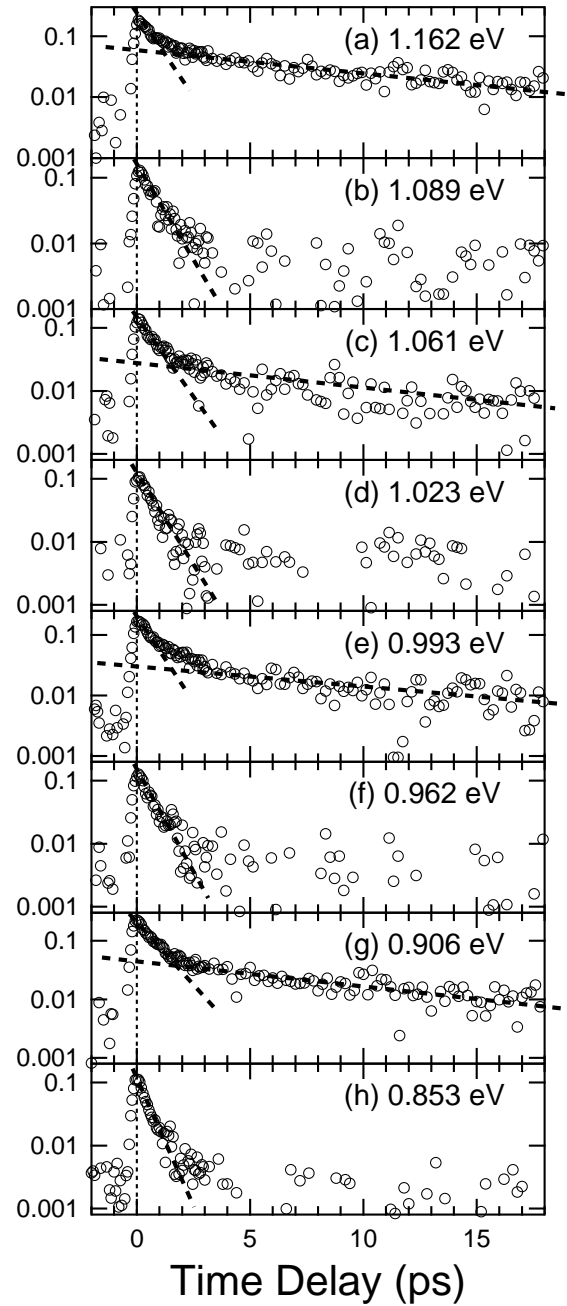}
}
\caption{Wavelength-dependent (one-color) pump-probe spectroscopy data,
corresponding to the photon energies labeled \#2, \#3, \#4, \#6, \#7, \#8,
\#11, and \#13 in
Fig. \ref{summary}(a), covering both absorption peaks and valleys.
For the energies corresponding
to absorption peaks [(a), (c), (e), and (g)], the chirality indices
($n$,$m$) of the SWNTs probed at those energies are
(a): (10,2), (c): (12,1) and (8,6), (e): (10,3) and (10,5), and (g):
(11,4).}
\label{pp-main}
\end{center}
\end{figure}

To study any differences between resonant and non-resonant excitations in carrier
relaxation as well as any diameter- and/or chirality-dependent phenomena,
we scanned the photon energy from
0.8 eV to 1.1 eV, corresponding to the E$_1$H$_1$ transitions of 0.82-1.29 nm
diameter tubes \cite{BachiloetAl02Science}.  For all the photon energies, the pump
fluence was kept constant at 1 mJ/cm$^{2}$, which is below the
saturation regime [see Fig. \ref{saturation}].
Figure \ref{summary}(a) shows the linear absorption
spectrum in the E$_1$H$_1$ transition range.
The photon energies at which we performed
pump-probe measurements are labeled 1 $-$ 16, covering both peaks and
valleys in absorption.  Figure
\ref{summary}(b) shows the maximum value of $\Delta T /T_0$ as a function of
photon energy; it loosely follows the absorption curve in (a).
Also shown in Fig. \ref{summary}(b)
(right vertical axis) is the ratio of the slow component
(defined as $\Delta T/T_0$ at 5 ps)
to the fast component (defined as $\Delta T/T_0$ at 0 ps) as a function
of photon energy; it also follows the absorption curve in (a), indicating
that the slow component is resonantly enhanced at absorption peaks.

To demonstrate this more directly,
eight traces of differential transmission dynamics taken at different
photon energies are shown in Figs.
\ref{pp-main}(a)-\ref{pp-main}(h).  The chosen photon energies correspond to the peaks
and valleys in the linear absorbtion data in Fig. \ref{summary}(a),
marked as 2, 3, 4, 6, 7, 8, 11, and 13.  For the photon energies corresponding
to peaks in linear absorption [(a), (c), (e), and (g)], the chirality indices
($n$,$m$), assigned through PL excitation spectroscopy
\cite{BachiloetAl02Science,WeismanBachilo03NL},
are (a): (10,2), (c): (12,1) and (8,6), (e): (10,3) and (10,5), and (g):
(11,4).
The slow component is clearly observable for the photon
energies corresponding to absorption peaks [(a), (c), (e) and (g)]
while the traces corresponding to valleys [(b), (d), (f) and (h)]
show only the initial, fast decay.

\subsection{pH Dependence}
\label{pH-dep}

\begin{figure}
\begin{center}
\resizebox{0.47\textwidth}{!}{%
\includegraphics{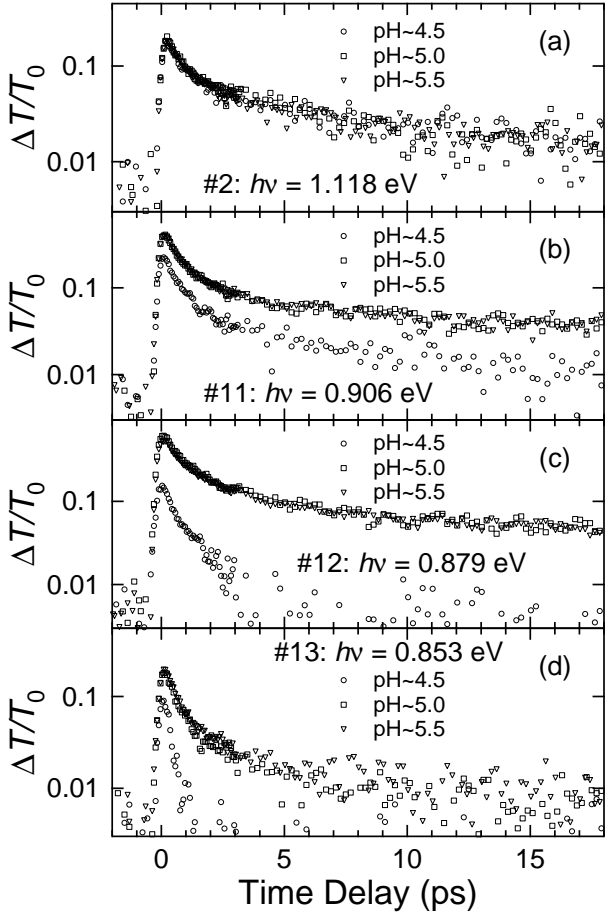}
}
\caption{pH-dependent pump-probe data for different wavelengths.
The pH dependence becomes stronger for smaller photon energies, corresponding
to larger-diameter SWNTs.}
\label{pH}
\end{center}
\end{figure}

\begin{figure}
\begin{center}
\resizebox{0.47\textwidth}{!}{%
\includegraphics{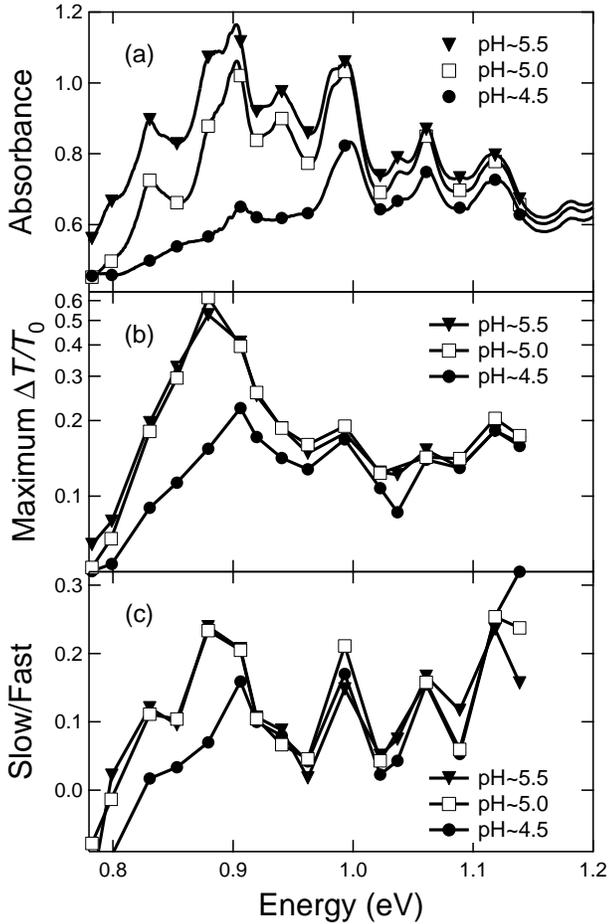}
}
\caption{pH-dependence of (a) near-infrared absorbance in the E$_1$H$_1$ energy
range, (b) maximum value of the differential transmission, and (c)
the ratio of the slow component ($\Delta T/T_0$ at 3 ps)
to the fast component ($\Delta T/T_0$ at 0 ps) as a function
of photon energy.}
\label{pH-abs}
\end{center}
\end{figure}

By adjusting the pH of the solution via adding NaOH or HCl,
we found that pump-probe dynamics are strongly dependent on the pH
and the dependence is stronger at longer wavelengths (or larger
tube diameters).  Specifically, we observed that the slow component
drastically diminishes as the pH is reduced.  Examples are shown in
Fig. \ref{pH}.  As previously reported \cite{StranoetAl03JPCB}, adding hydrogen ions,
H$^+$,
to the solution (or, equivalently, decreasing the pH value) diminishes,
and finally collapses, linear absorption and PL peaks.  This effect starts from the
longer wavelength side, i.e., from larger diameter tubes.
The corresponding
reduction and disappearance of the slow component shows exactly the same trend,
as shown in Fig. \ref{pH}.  For example, at $h \nu$ = 1.118 eV (or 1.11 $\mu$m), there is
almost  no change in pump-probe dynamics from pH = 5.5 to pH = 4.5 [see
Fig. \ref{pH}(a)].  However, as the photon energy is decreased to 0.906 eV [(b)],
0.879 eV [(c)], and 0.853 eV [(d)], the disappearance of the slow component
becomes increasingly more drastic.
Figures \ref{pH-abs}(a), \ref{pH-abs}(b), and \ref{pH-abs}(c) show, respectively,
the pH-dependence of (a) near-infrared absorbance spectra in the E$_1$H$_1$ energy
range, (b) the maximum value of the differential transmission, and (c)
the ratio of the slow component ($\Delta T/T_0$ at 3 ps)
to the fast component ($\Delta T/T_0$ at 0 ps) as functions
of photon energy.   It is clearly seen that all the three quantities decrease
with decreasing pH, and the decrease is more significant at longer wavelengths
(or wider-diameter tubes).

\section{Discussion}
\label{discussion}

As we presented in the last section, our wavelength-dependent
single-color pump-probe spectroscopy data on micelle-suspended SWNTs revealed
two different decay components, suggesting two different carrier relaxation
processes.  One component is fast (0.3-1.2 ps) and the other is slow (5-20 ps).
In the following we discuss possible origins of these two dynamical processes as well as
the intriguing pH dependence of the pump-probe data we obtained.

Both the positive sign of the differential transmission and the initial ultrafast
(< 1 ps) relaxation agree with the recent reports for semiconducting SWNTs
\cite{ChenetAl02APL,HanetAl02APL,LauretetAl03PRL}.
The positive sign can be interpreted as state filling as the cause of the
photo-bleaching signal and is consistent with
the saturation at high fluences (Fig. \ref{saturation}).
Namely, the pump-induced carriers fill states and reduce the probe absorption.
We believe that the ultrafast decay mechanism is {\em non-radiative}
(i.e., phonon and/or impurity-mediated) and {\em intraband}.
It is likely to be non-radiative because this signal exists even for samples
that do not luminesce \cite{ChenetAl02APL,HanetAl02APL,LauretetAl03PRL}.
It is likely to be intraband relaxation (as opposed to interband relaxation)
since it exists both in resonant and non-resonant cases.
Note that even when the photon energy is resonant
with the E$_1$H$_1$ absorption peak of a particular tube type, it creates non-resonant
carriers in other types of tubes  whose band gaps are smaller than that
of the resonant tube.
Note also that we are not seeing any pump-power-dependent decay times,
as shown in Fig. \ref{power}, which
precludes carrier-density-dependent non-radiative processes such as
Auger recombination as the origin of the fast decay component.
In addition, it is likely that there is some coherent contribution to the pump-probe
signal in this ultrashort time scale.  However, since we have not observed any
four-wave mixing signal, we do not have an estimate on the dephasing time and thus
will not discuss how large this contribution is.

On the other hand, the slow decay signal, which was resonantly enhanced when the
photon energy coincided with an interband absorption peak,
has not been reported previously.
We interpret this slow component as interband carrier recombination.
Note that in bundled samples band edges are not well defined for individual tubes
due to electronic coupling, and thus photo-created carriers cannot show
clear interband dynamics in pump-probe measurements.
In other words, there is no clear distinction between inter-tube dynamics and
intraband dynamics in bundled tubes because van Hove singularities
of different tube types form a van Hove absorption {\em band}.
We further argue that the slow component we observed is probably related to radiative
recombination.
An important fact for supporting this claim is that in
the previous work no PL was observed whereas our sample exhibits
PL peaks at the same energies as E$_1$H$_1$ absorption peaks (see
Fig. \ref{linear}).  In addition,
reducing the pH of the solution destroys PL and absorption peaks while at
the same time the slow decay also vanishes (see Fig. \ref{pH}),
indicating an {\em intimate
relationship between PL and the slow decay signal}.
The actual value of the radiative recombination lifetime $\tau_r$ depends on
the value of the radiative efficiency (or quantum yield)
 $\eta$ = $\tau_r^{-1}/(\tau_r^{-1} + \tau_{nr}^{-1})$,
where $\tau_{nr}$ is the nonradiative lifetime, since what we measure
experimentally is the {\it total} decay rate, $\tau^{-1} = \tau_r^{-1} + \tau_{nr}^{-1}$.
Direct measurements of $\tau_r$ through time-resolved PL are in progress.

Finally, we propose possible scenarios for the drastic pH dependence we observed.
Decreasing the pH leads to an increase in the density
of H$^+$.  First, adsorbed H$^+$ ions on the nanotube surface
could add a relaxation channel by the creation of ultrafast
carrier trapping centers (defects).
This should make the fast non-radiative recombination dominant
over the slower radiative recombination if the density of such trapping centers
is high.
In addition, as previously noted, H$^+$ ions effectively act as acceptors in SWNTs,
and thus the Fermi energy ($E_F$) depends on the H$^+$ density.
When the density is so high that $E_F$ lies inside the valence band,
interband absorption peaks disappear, irrespective of whether the peaks are due
to excitons or van Hove singularities.
This can be viewed as a 1-D manifestation of the well-known
Burstein-Moss effect \cite{Burstein54PR,Moss54PPS}.
Larger-diameter tubes are expected to have smaller acceptor binding energies due to
their smaller effective masses, and thus should be more susceptible to pH changes.
In a heavy doping regime,
there is a degenerate hole gas in the valence band, which can completely quench
excitonic processes.

\section{Summary}
\label{summary-future}

In summary, we have carried out an ultrafast optical
study of micelle-suspended single-walled carbon nanotubes.
We have observed two relaxation
regimes in the dynamics of micelle-suspended
single-walled carbon nanotubes under resonant excitations.
We interpret the previously observed shorter decay time $\tau$$_{1}$
as nonradiative intraband relaxation and the previously unobserved larger decay
time $\tau$$_{2}$ as related to radiative interband recombination.
The relaxation of photoexcited carriers can be made faster
by increasing the density of H$^+$ ions in the solution while simply
increasing the number of photoexcited carriers does not change the
dynamics.  The sensitive pH dependence provides
a novel means to chemically control carrier states and dynamics
in nanotubes.

\section{Acknowledgements}
\label{thanks}
This work was supported by the
Robert A. Welch Foundation (Grant No. C-1509), the Texas Advanced
Technology Program (Project No. 003604-0001-2001), and the
National Science Foundation CAREER Award (Grant No. DMR-0134058).


\end{document}